\documentclass[12pt]{article}
\usepackage{amsfonts}
\usepackage{graphicx}
\usepackage{amsmath}
\usepackage{float}
\textwidth=6.5in \hoffset=-.75in \voffset=-1in \numberwithin{equation}{section}
\numberwithin{figure}{section}

\newcommand {\nn}{\nonumber}
\newcommand {\be}{\begin{equation}}
\newcommand {\ee}{\end{equation}}
\begin{document}

\begin{titlepage}
\vspace{1cm}
\begin{center}
{\Large \bf {Exact Solutions to Einstein-Maxwell theory on Eguchi-Hanson Space}}\\
\end{center}
\vspace{2cm}
\begin{center}
A. M. Ghezelbash{ \footnote{ E-Mail: masoud.ghezelbash@usask.ca}}, V. Kumar{ \footnote{ E-Mail: v.kumar@usask.ca }}
\\
Department of Physics and Engineering Physics, \\ University of Saskatchewan, \\
Saskatoon, Saskatchewan S7N 5E2, Canada\\
\vspace{1cm}
\vspace{2cm}
%\today\\
\end{center}

\begin{abstract}
In this article, we construct explicit analytical exact solutions to the six and higher dimensional Einstein-Maxwell theory. In all solutions, a subspace of the metric is the Eguchi-Hanson space where the metric functions are completely determined in terms of known analytical functions. Moreover, we find the solutions can be extended to non-stationary exact solutions to Einstein-Maxwell theory with cosmological constant.  We show that the solutions are asymptotically expanding patches of de-Sitter spacetime.

\end{abstract}
\end{titlepage}\onecolumn 
\bigskip 

\section{Introduction}
\label{sec:Intro}
%
%\[
%Later: ShortLiterature Review, OldConvSolutionsForTN, OldSolutionsWithTNWithMoreMatterFields \]
%\[
%ImportanceOrganization
%\]
With the recent experimental confirmation of gravitational waves \cite{GravWave}, there is a renewed interest in gravitational physics and general relativity, which has so far defied any attempt to unification with quantum theory. Many approaches to quantum gravity involve more than four spacetime dimensions and as such, it is important to explore exact solutions in general relativity in higher dimensions.

The black hole solutions, as well as the other solutions to the Einstein-Maxwell theory in different dimensions, provide interesting results and better understanding of physical processes such as coalescing of black holes \cite{Co1}-\cite{Co3} or possible non-spherical horizon topologies for the higher-dimensional black holes \cite{Em1}-\cite{El1}.  

%s also attracted the interest of experimentalists using particle colliders to table-top experiments.

Moreover, the other type of solutions to Einstein-Maxwell theory, include convoluted-like solutions \cite{TN}, Nutty solutions \cite{awad}, solitonic and dyonic solutions \cite{EY21}-\cite{EY2222} and also braneworld solutions \cite{EY41}-\cite{EY43}. In \cite{EY51}, the authors found the solutions of Einstein's equations in the presence of a scalar field with a Liouville potential, that is coupled to the Maxwell field. The string theory extension of Einstein-Maxwell theory was considered in \cite{EY5string1}.  %We coand string theory extended solutions \cite{EY5}.
Moreover, in \cite{me}, the author found exact solutions to five-dimensional cosmological Einstein-Maxwell theory coupled to a dilaton field in which there are two coupling terms; one coupling between the dilaton and electromagnetic field strength and the other coupling between the dilaton and cosmological constant. The solutions to extension of Einstein-Maxwell theory that include axion field and also Chern-Simons term as well as supergravity solutions, were explored extensively in \cite{other13}-\cite{other16}.

%Moreover, solutions The supegravity solutions and cosmological solutions and solutions with dilaton, axion and Chern-Simons term are considered in \cite{other13}-\cite{}. 
The motivation for this article is based on the recent discovered solutions in the eleven dimensional supergravity \cite{M1}-\cite{M3} where the authors constructed new convoluted-like solutions to the low energy limit of M-theory. These solutions upon compactification over a compact dimension of transverse geometry provide the M-theory realization of fully localized intersecting branes in type IIA superstring theory. The solutions are based on self-dual transverse geometries \cite{M1}-\cite{M3} and possess eight supercharges. A key feature of the solutions is that the metric function for the M-branes can be expressed analytically in terms of convoluted-like intergrals which in fact, is a result of separability of the field equations for the special form of the metric and gauge field ansatzes.  Inspired by the convoluted-like membrane solutions based on self-dual transverse geometries as well as convoluted-like solutions to Einstein-Maxwell theory based on Taub-NUT space \cite{TN}, in this article,  we construct exact analytical convoluted-like solutions to Einstein-Maxwell theory (with and without cosmological constant) in six and higher dimensions. The minimal dimension of six for the theory, is required to have nontrivial convoluted-like solutions.
%of  hat provide the realization of fully localized type IIA D2-branes intersecting we constuct explicit analytical exact solutions to the Einstein-Maxwell and also Einstein-Maxwell-Lambda theories in six and higher dimensions. 

This article is organized as follows. In section \ref{sec:EH}, we review the Eguchi-Hanson instantons and their properties. In section \ref{sec:DD},
we derive the exact convoluted-like solutions to the Einstein-Maxwell theory in any dimensions greater than or equal to six. We find that, in general, the solutions can be written as integrals that include two unknown functions. These functions depend on separation constant, which appear in separated equations of motions. We find and fix these functions, by solving some integral equations. The integral equations guarantee the exact solutions are reducing properly to some known solutions, in appropriate limits of some parameters in the theory. % the and as a result,  We then fix the measure functions in the convoluted integral by comparing the appropriate limits of the solutions to that of exact flat solutions to the Einstein-Maxwell theory. 
Moreover, in section \ref{sec:cosmo}, we find the exact analytical solutions to Einstein-Maxwell-Lambda theory and discuss some features of the solutions. 
%present the convoluted solutions for the seven dimensional theory. Inspired by the solutions found in section \ref{sec:6D}, we construct the exact convoluted solutions for the $D$-dimensional Einstein-Maxwell theory in section \ref{sec:DD}.
%In section \ref{sec:2ndsolutions}, we find the second set of solutions in $D$-dimensions by analytically continuing the separation constant.
%In section \ref{sec:cosmo}, we consider the Einstein-Maxwell theory with positive cosmological constant and show the equations of motion are separable if one consider a proper separation of three coordinates in the metric functions. 
We present the concluding remarks in section \ref{sec:con}, followed by an appendix.

\section{The geometry of Eguchi Hanson instantons}
\label{sec:EH}
%\[
%MaybeMergeThisSectionIntoIntroduction?
%\]
The self-dual Eguchi-Hanson metric was discovered by Eguchi and Hanson \cite{EH78} and independently by Calabi \cite{Cal79} as a gravitational analogue to the finite action BPST instanton solution in Yang-Mills theory. Consequently, it is generally regarded as a gravitational instanton, which may be defined as a complete non-singular positive-definite solution to Einstein's equations with or without the cosmological constant.%$\Lambda$ term.
 These ``pseudoparticles" play an important role in Euclidean quantum gravity and string theoretical approaches to quantum gravity. In Yang-Mills theory, they describe the tunnelling amplitudes between vacua and  give the dominant contribution to the path integral.

The four-dimensional Eguchi-Hanson space is a non-compact and self-dual solution to Einstein's equations, where the metric defined over the cotangent bundle $T^*\mathbb{P}^1(\mathbb{C})$ (or $T^*S^2$), is given by the line element
%The line element for the Eguchi-Hanson space is represented by
\be
ds_{EH}^2=\frac{dr^2}{g(r)^2}+\frac{r^2}{4}(\sigma_1^2+\sigma_2^2)+\frac{r^2g(r)^2}{4}\sigma_3^2,\label{EH}
\ee
where the metric function $g(r)=\sqrt{1-(\frac{a}{r})^4}$ and $\sigma_i,i=1,2,3$ are the $SU(2)$ group manifold left invariant one forms
\begin{eqnarray}
\sigma_1&=&\sin \psi d\theta-\cos \psi\sin \theta d \phi,\\
\sigma_2&=&-\cos \psi d\theta-\sin \psi\sin \theta d \phi,\\
\sigma_3&=&d\psi +\cos \theta d \phi.
\end{eqnarray}
The $SU(2)$ group manifold is parameterized by three periodic Euler angles $\theta, \phi$ and $\psi$ with periodicities of $\pi, 2\pi$ and $4\pi$ respectively and the radial coordinate $r>a$. However the smoothness of the Eguchi-Hanson space for any positive Eguchi-Hanson parameter $a$ as well as removing the bolt singularity at $r=a$ implies that the coordinate $\psi$ has a period of $2\pi$. The Euler characteristic $\chi$ is 2 (same as $S^2$), with the Hirzebruch signature $\tau$ equal to -1 and the Dirac spin-$\frac{1}{2}$ index 0. \cite{EH79} .The boundary at $r\rightarrow\infty$ is the lense space $S^3/\mathbb{Z}_2$ (or $\mathbb{P}^3(\mathbb{R})$) and so the Eguchi-Hanson space is asymptotically locally Euclidean.
The Eguchi-Hanson metric possesses a self-dual Kahler two form and the metric at $r=a$ describes a $S^2$ with radius $\frac{a}{2}$. 
%The Eguchi-Hanson space has been used to construct warped solutions and can be used to . 
The warped extension of Eguchi-Hanson geometry has been used to construct and study the CFT realization of the heterotic strings in the double scaling limit \cite{HetWarp}. The Eguchi-Hanson like solutions in extended theories of gravity have also been investigated in \cite{Rob,Rob2}. Moreover, the coalescence of black holes on the Eguchi-Hanson geometry was extensively studied in \cite{Co2}, \cite{Co3}, \cite{CBH2},  \cite{CBH4}. Some black ring and brane solutions also are constructed and studied in \cite{BR1}-\cite{BR3}.

\section{Convoluted-like solutions in $D\geq 6$ Einstein-Maxwell theory}
\label{sec:DD}

\begin{figure}[H]%HAVE BOTH .eps and converted to PDF in DIRECTORY
\centering
\includegraphics[width=0.4\textwidth]{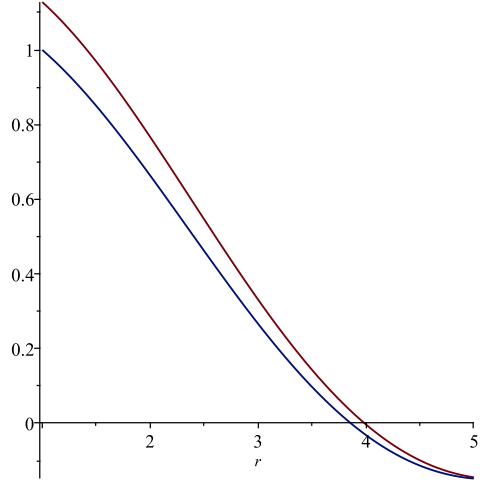}%,\includegraphics[width=0.5\textwidth]{dil}

\caption{The ${H_C} \left( 0,0,0,-\frac{{
a}^{2}{c}^{2}}{2},\frac{{a}^{2}{c}^{2}}{4},{\frac {{a}^{2}-{r}^{2}}{2{a}^{
2}}} \right)$ function that appear in solutions (\ref{HFIN}), as function of $r$, where we set $a=1,c=1$ (blue) and $a=1,c=2$ (red).}
\label{fig1}
\end{figure}

We consider the following 
$D$-dimensional metric ansatz \cite{TN}
\be
ds_{D}^{2}=-H(r,x)^{-2}dt^{2}+H(r,x)^{2/(D-3)}(dx^2+x^2d\Omega_{D-6}+ds_{EH}^2),\label{mD}
\ee
where $d\Omega_{D-6}$ is the metric on unit sphere $S^{D-6}$.  We only consider the solutions with electric field that the only non-zero component of the gauge field is given by
\be
{A_t}={\sqrt{\frac{D-2}{D-3}}H^{-1}(r,x)}\label{gaugeD}.
\ee 
The ansatzes (\ref{mD}) and (\ref{gaugeD}) for the $D$-dimensional metric and the gauge field were considered previously to construct the convoluted-like solutions in Einstein-Maxwell theory based on a four dimensional Taub-NUT instanton \cite{TN}.
The Einstein's equations in presence of electromagnetic field together with the Maxwell's equations are satisfied, provided that $H(r,x)$ satisfies a second order partial differential equation. The differential equation for $H(r,x)$ is completely separable in any dimensions, if we set $H(r,x)=\alpha+\beta R(r) X(x)$ where $\alpha$ and $\beta$ are two constants. Without losing any generality, we set $\alpha=1$. We find that after separation of the coordinates, the differential equation for $R(r)$ is given by,
\be
{\frac {d^{2}}{d{r}^{2}}}{R} \left( r \right) +{\frac {   \left( {a}^{4}+3\,{r
}^{4} \right) }{r \left( r^4-{a}^{4} \right) }} {
\frac {d}{dr}}{R} \left( r \right)+{\frac {{\it R}
 \left( r \right) {r}^{4}{c}^{2}}{r^4-{a}^{4}}}=0,\label{Rde}
\ee
where $a$ is the Eguchi-Hanson parameter in Eguchi-Hanson metric (\ref{EH}) and we show the separation constant by $c^2$.
The solutions to differential equation (\ref{Rde}) are
\begin{eqnarray}
{\it R} \left( r \right) &=&{r_1}\,{ H_C} \left( 0,0,0,-\frac{{
a}^{2}{c}^{2}}{2},\frac{{a}^{2}{c}^{2}}{4},{\frac {{a}^{2}-{r}^{2}}{2{a}^{
2}}} \right)+ \nn\\
&+&{r_2}\,{ H_C} \left( 0,0,0,-\frac{{
a}^{2}{c}^{2}}{2},\frac{{a}^{2}{c}^{2}}{4},{\frac {{a}^{2}-{r}^{2}}{2{a}^{
2}}} \right) 
\int \! \frac{f(r)dr}{\left(     { H_C} \left( 0,0,0,-\frac{{
a}^{2}{c}^{2}}{2},\frac{{a}^{2}{c}^{2}}{4},{\frac {{a}^{2}-{r}^{2}}{2{a}^{
2}}} \right) 
     \right) ^{2}},\nn\\
&&\label{Rsol}
\end{eqnarray}
where $r_1$ and $r_2$ are two constants. The function $f(r)$ is given by
\be
f(r)=\frac{r}{r^4-a^4},
\ee
and $H_C$ is the Heun-C function. The Heun-C function $H_C(\alpha,\beta,\gamma,\delta,\rho,z)$ represents the solutions to the confluent second order differential Heun equation 
\begin{eqnarray}
&&\frac{d^2\,H_C}{dz^2}+\frac{(\alpha z^2+(\beta+\gamma+2-\alpha)z-\beta-1}{z(z-1)}\frac{d\,H_C}{dz}+\nonumber\\
&+&\frac{\{[(\beta+\gamma+2)\alpha+2\delta]z-(\beta+1)\alpha+(\gamma+1)\beta+2\rho+\gamma\}}{2z(z-1)}H_C=0,
\end{eqnarray}
with the boundary conditions $H_C(\alpha,\beta,\gamma,\delta,\rho,0)=1$ and $\frac{d}{dz}H_C(\alpha,\beta,\gamma,\delta,\rho,z)\vert_{z=0}=\frac{(1+\gamma-\alpha)\beta+\gamma+2\rho-\alpha}{2(\beta+1)}$.
The confluent Heun equation is obtained from the general Heun equation through a confluence process in which two singularities coalesce into one.  We note that Heun equations have appeared in several works on curved spacetimes, including those involving the Eguchi-Hanson space \cite{Mal03}.
Moreover, we find that the differential equation for $X(x)$ is

\be
x{\frac {d^{2}}{d{x}^{2}}}{X} \left( x \right) + \left( D-6
\right) {\frac {d}{dx}}{X} \left( x \right) -c^2 x {X} \left( x
\right) =0. \label{XDeq}
\ee
We find that the solutions to (\ref{XDeq}) are given in terms of Bessel functions $I$ and $K$ by
\be
{X} \left( x \right) ={x}^{\frac{7-D}{2}}(
x_1\,{{I_{\frac{D-7}{2}}}\left(cx\right)}+x_2\,
{{K_{\frac{D-7}{2}}}\left(cx\right)}),\label{Xsol}
\ee
where $x_1$ and $x_2$ are constants.
We can then superimpose all the solutions (\ref{Rsol}) and (\ref{Xsol}) with different values of separation constant $c$, to find the most general solution to Einstein's equations in presence of gauge field (\ref{gaugeD}).  We note that both the solutions (\ref{Xsol}) and (\ref{Rsol}) are real-valued functions.%, however the solutions  to the differential equation (\ref{Rde}) are not real-valued functions.  As we notice, the different functions that appear in the differential equation (\ref{Rde}) are real and so the real part of the solutions (\ref{Rsol}), also is a solution to the differential equation (\ref{Rsol}).  
To simplify our analysis, we consider $r_2=0$ in equation (\ref{Rsol}). As a result, % and consider the real part of Heun-C function, to 
we find a general solution for the metric function $H(r,x)$, which is given by 
\begin{eqnarray}
H(r,x)&=&1+ \int _{0}^\infty dc\, {x}^{\frac{7-D}{2}}\,{  H_C} \left( 0,0,0,-\frac{{
a}^{2}{c}^{2}}{2},\frac{{a}^{2}{c}^{2}}{4},{\frac {{a}^{2}-{r}^{2}}{2{a}^{
2}}} \right)\{{h_{1}(c)}\,
{{I_{\frac{D-7}{2}}}\left(cx\right)}\nonumber\\
&+&{h_{2}(c)}
{{K_{\frac{D-7}{2}}}\left(cx\right)}\} \label{HGEH},
\end{eqnarray}
where %$\bar H_C =\Re  H_C $ and 
we include two measure functions $h_1(c)$ and $h_2(c)$ in the integrand, that are functions of separation constant $c$. Figure \ref{fig1} shows the typical behaviour of ${ H}_C$ as function of $r$, as a part of integrand in the solution (\ref{HGEH}), where we set $a=1$ and $c=1,2$.
%\[
%AddReasonWhyWeDontConiderTheSecondSolutionInAboveIntegral. 
%\]
%We note that we include  as two arbitrary measure functions in the solutions. 
To fix the measure functions in the general solution (\ref{HGEH}), we notice that in the limit $a \rightarrow 0$ (or equivalently as $r \rightarrow \infty$, since the Eguchi-Hanson metric function $g(r)$ in (\ref{EH}) approaches to 1), the four-dimensional Eguchi-Hanson (\ref{EH}) is asymptotically $R^4$ with the line element $dr^2+r^2d\Omega_3^2$. 

Moreover, we find the following exact solutions to the Einstein-Maxwell theory, where the 
$D$-dimensional metric is given by

\be
ds_0^2=-\frac{1}{H_{0}^2(r,x)}dt^2+H_{0}(r,x)^{2/(D-3)}(dx^2+x^2d\Omega_{D-6}+dr^2+r^2d\Omega_3^2),\label{Ehazero}
\ee
%\[
%Double CheckThis   DONE
%\]
and the gauge field is given by
\be
{A_t}={\sqrt{\frac{D-2}{D-3}}H_0^{-1}(r,x)}\label{gaugeD0},
\ee
The metric function $H_0(r,x)$ in $D$-dimensions is given analytically by
\be
H_{0}(r,x)=1+{\frac {\chi}{ \left( {r}^{2}+{x}^{2} \right) ^{\frac{D-3}{2}}}}\label{H09},
\ee
where $\chi$ is a constant. To find and fix the measure functions that appear in (\ref{HGEH}), we notice that, in the limit of $a\rightarrow 0$ (or  $r\rightarrow \infty$), the metric function (\ref{HGEH}) for the spacetime (\ref{mD}), must approach to the metric function (\ref{H09}) for the spacetime (\ref{Ehazero}), since in the limit  of $a\rightarrow 0$, the spacetime (\ref{mD}) reduces to (\ref{Ehazero}). This yields an integral equation for the measure functions $h_1(c)$ and $h_2(c)$
\begin{equation}
\int _{0}^\infty dc\, {x}^{\frac{7-D}{2}}\,\lim _{a\rightarrow 0}{  H_C} \left( 0,0,0,-\frac{{
a}^{2}{c}^{2}}{2},\frac{{a}^{2}{c}^{2}}{4},{\frac {{a}^{2}-{r}^{2}}{2{a}^{
2}}} \right)\{{h_{1}(c)}\,
{{I_{\frac{D-7}{2}}}\left(cx\right)}
+{h_{2}(c)}
{{K_{\frac{D-7}{2}}}\left(cx\right)}\} ={\frac {\chi}{ \left( {r}^{2}+{x}^{2} \right) ^{\frac{D-3}{2}}}}\label{inteqD}.
\end{equation} 
We also note that equation (\ref{inteqD}) guarantees the gauge field (\ref{gaugeD}) also approaches the gauge field (\ref{gaugeD0}), in the limit of $a\rightarrow 0$ (or  $r\rightarrow \infty$).
By solving the equation (\ref{Rde}) in the limit $a\rightarrow 0$ and comparing its solution to (\ref{Rsol}), we find that
\begin{equation}
\lim _{a\rightarrow 0}{ H_C} \left( 0,0,0,-\frac{{
a}^{2}{c}^{2}}{2},\frac{{a}^{2}{c}^{2}}{4},{\frac {{a}^{2}-{r}^{2}}{2{a}^{
2}}} \right)=\frac{2}{cr}J_{1}\left(cr\right) \label{HCBJ},
\end{equation}
where $J_{1}$ is a Bessel function of the first kind of $r$.  %We also find that 
%\[
%Okay,SeemsALittleBitOdd!MoreInfoForLimitOfHC
%\]
We also find the following set of integrals %in different dimensions $D$
\begin{equation}
\int _0^\infty dc\,c^{\frac{D-3}{2}}\,\frac{J_1(cr)}{r}\,x^{-\frac{D-7}{2}}\,K_{\frac{D-7}{2}}(cx) = \frac{\xi_D}{\left( {r}^{2}+{x}^{2} \right) ^{\frac{D-3}{2}}},\label{JKint}
\end{equation}
where the constant $\xi_D$ depends on dimension $D$. We find that for even dimensions %different values of the constant as
%\be
%\xi_6=\sqrt{\frac{\pi}{2}},\xi_7=2,\xi_8=3\sqrt{\frac{\pi}{2}},\xi_9=8,\xi_{10}=15\sqrt{\frac{\pi}{2}},\xi_{11}=48,\xi_{12}=105\sqrt{\frac{\pi}{2}},%\cdots
\be
\xi_{6+2n}=\sqrt{\frac{\pi}{2}}(2n+1)!!,
\ee
%\ee
and for odd dimensions
\be
\xi_{7+2n}=(2n+2)!!,
\ee
where $n=0,1,\cdots$. 
So, we find that the measure functions $h_1(c)$ and $h_2(c)$ in $D$-dimensions, are given by
\be
h_1(c)=0,\,h_2(c)=\frac{\chi}{2\xi_D}c^{\frac{D-1}{2}},
\ee
and so the most general solution for the metric function $H(r,x)$ in the metric (\ref{mD}) and the gauge field (\ref{gaugeD}), is given by
%\begin{eqnarray}
\be
H(r,x)=1+ \frac{\chi}{2\xi_D}\int _{0}^\infty dc\, {x}^{\frac{7-D}{2}}\,c^{\frac{D-1}{2}}{H_C} \left( 0,0,0,-\frac{{
a}^{2}{c}^{2}}{2},\frac{{a}^{2}{c}^{2}}{4},{\frac {{a}^{2}-{r}^{2}}{2{a}^{
2}}} \right)
{{K_{\frac{D-7}{2}}}\left(cx\right)} \label{HFIN}.
%\end{eqnarray}
\ee

Moreover, we can find another independent exact solution for the metric function by analytically continuing the separation constant $c$ in equations (\ref{Rde}) and (\ref{XDeq}) to $ic$. The solutions to equation (\ref{Rde}) with $c$ replaced by $ic$ are given by the Heun-C functions $H_C(0,0,0,\frac{a^2c^2}{2},
-\frac{a^2c^2}{4},\frac{a^2-r^2}{2a^2})$. Moreover the solutions to equation (\ref{XDeq}) with $c$ replaced by $ic$, are given by as linear combination of the Bessel functions 
${{J_{\frac{D-7}{2}}}\left(cx\right)}$ and
$
{{Y_{\frac{D-7}{2}}}\left(cx\right)}
$. So, we consider the superposition of the solutions as the most general solution for the metric function which is given by
\begin{eqnarray}
\tilde H(r,x)&=&1+ \int _{0}^\infty dc\, {x}^{\frac{7-D}{2}}\,{{H}_C} \left( 0,0,0,\frac{{
a}^{2}{c}^{2}}{2},-\frac{{a}^{2}{c}^{2}}{4},{\frac {{a}^{2}-{r}^{2}}{2{a}^{
2}}} \right)\{{\tilde h_{1}(c)}\,
{{J_{\frac{D-7}{2}}}\left(cx\right)}\nonumber\\
&+&{\tilde h_{2}(c)}
{{Y_{\frac{D-7}{2}}}\left(cx\right)}\} \label{HGEH2},
\end{eqnarray}
We find that the limit of Heun-C function in the limit of $a\rightarrow 0$ is given by
\begin{equation}
\lim _{a\rightarrow 0}{ { H}_C} \left( 0,0,0,\frac{{
a}^{2}{c}^{2}}{2},-\frac{{a}^{2}{c}^{2}}{4},{\frac {{a}^{2}-{r}^{2}}{2{a}^{
2}}} \right)=\frac{2}{cr}I_{1}\left(cr\right) \label{HCBJ},
\end{equation}
and so we find the following integral equation for the measure functions
\begin{equation}
\int _{0}^\infty dc\,c^{-1} {x}^{\frac{7-D}{2}}\,\frac{I_1(cr)}{r}\{{\tilde h_{1}(c)}\,
{{J_{\frac{D-7}{2}}}\left(cx\right)}
+{\tilde h_{2}(c)}
{{Y_{\frac{D-7}{2}}}\left(cx\right)}\}={\frac {\chi}{ 2\left( {r}^{2}+{x}^{2} \right) ^{\frac{D-3}{2}}}}\label{ie2}.
\end{equation}
To find the solutions to the integral equation (\ref{ie2}), we consider the analytically continuations $r\rightarrow ir$ and $x\rightarrow ix$ in equation (\ref{JKint}) an find the following integrals
\begin{equation}
\int _0^\infty dc\,c^{\frac{D-3}{2}}\,\frac{I_1(cr)}{r}\,x^{-\frac{D-7}{2}}\{(-1)^{\frac{D-7}{2}}\,J_{\frac{D-7}{2}}(cx)+iY_ {\frac{D-7}{2}}(-cx)= -\frac{2(-1)^{\frac{3-D}{2}}}{\pi }\frac{\xi_D}{\left( {r}^{2}+{x}^{2} \right) ^{\frac{D-3}{2}}}\label{myint}.
\end{equation}
As a result of comparing (\ref{ie2}) with (\ref{myint}), we find that the only consistent measure functions for even $D=6,8,\cdots$ are 
\be
\tilde h_1(c)=\frac{c^{\frac{D-1}{2}}\chi \pi}{4\xi_D},\,\tilde h_2(c)=0,
\ee
where we should consider $x>0$. 
Moreover for odd $D=7,9,\cdots$, we find that the only consistent measure functions are given by
\begin{equation}
\tilde h_1(c)=-\frac{c^{\frac{D-1}{2}}\chi \pi}{4\xi_D},\,\tilde h_2(c)=0,
\ee
where we should consider $x<0$.
So, we find the second general solution as
\begin{equation}
\tilde H(r,x)=1+ \frac{\chi \pi  \epsilon_D}{4\xi_D}\int _{0}^\infty dc\, {c^{\frac{D-1}{2}}}\, {x}^{\frac{7-D}{2}}\,{ { H}_C} \left( 0,0,0,\frac{{a}^{2}{c}^{2}}{2},-\frac{{a}^{2}{c}^{2}}{4},{\frac {{a}^{2}-{r}^{2}}{2{a}^{
2}}} \right)\,
{{J_{\frac{D-7}{2}}}\left(cx\right)}
 \label{GENHGEH2},
\end{equation}
where $\epsilon_D$ is equal to $1$ for even dimensions and $-1$ for odd dimensions. Figure \ref{fig2} shows the typical behaviour of ${ H}_C$ as function of $r$, as a part of integrand in the general solution (\ref{GENHGEH2}), where we set $a=1$ and $c=1,2$. As we notice the Heun-C functions in (\ref{GENHGEH2}) are monotonically divergent at large $r$, however, the Bessel functions in  (\ref{GENHGEH2}), for large values of coordinate $x$ have oscillatory decaying behaviour to zero values.

\begin{figure}[H]%HAVE BOTH .eps and converted to PDF in DIRECTORY
\centering
\includegraphics[width=0.4\textwidth]{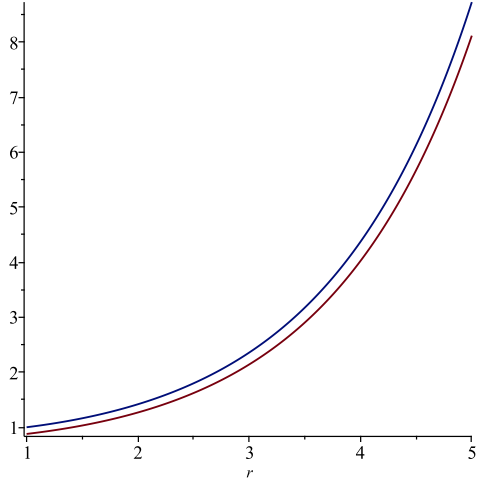}%,\includegraphics[width=0.5\textwidth]{dil}

\caption{The ${ { H}_C} \left( 0,0,0,\frac{{a}^{2}{c}^{2}}{2},-\frac{{a}^{2}{c}^{2}}{4},{\frac {{a}^{2}-{r}^{2}}{2{a}^{
2}}} \right)$ function that appear in the second solutions (\ref{GENHGEH2}), as function of $r$, where we set $a=1,c=1$ (blue) and $a=1,c=2$ (red).  }
\label{fig2}
\end{figure}

%\section{The second class of $D$-dimensional solutions}
%\label{sec:2ndsolutions}

\section{The cosmological convoluted-like solutions in $D\geq 6$ dimensional Einstein-Maxwell Theory}
\label{sec:cosmo}
In this section, we consider the Einstein-Maxwell theory in presence of positive cosmological constant. We  consider the metric ansatz as \cite{TN}
\be
ds_{D}^{2}=-H(t,r,x)^{-2}dt^{2}+H(t,r,x)^{2/(D-3)}(dx^2+x^2d\Omega_{D-6}+ds_{EH}^2),\label{mDtime}
\ee
Moreover, we consider the only non-zero component of the gauge field as
\be
{A_t}(t,r,x)={\sqrt{\frac{D-2}{D-3}}H^{-1}(t,r,x)}\label{gaugeDtime}.
\ee 
%
%the metric ansatz as 
%\be
%ds_6^{2}=-H(t,r,x)^{-2}dt^{2}+H(t,r,x)^{2/3}(dx^2+ds_{n}^2),\label{D6Cos}
%\ee
We should note that the Einstein and Maxwell field equations can be satisfied only if the metric function $H$ depends on time coordinate $t$, as well as the two spatial coordinates  $r,x$ (as in section \ref{sec:DD}) in presence of cosmological constant.
The Einstein's and Maxwell's equations, in presence of cosmological constant, lead to different partial differential equations for $H(t,r,x)$ in different dimensions (Appendix A). 
%the following second order partial differential equation for $H(t,r,x)$
%\begin{eqnarray}
%&&-2H^{11/3}V^3r^4\frac{\partial ^2 H}{\partial t^2}-\frac{10}{3}H^{8/3}V^3r^4(\frac{\partial H}{\partial t})^2+2V^3r^4H\frac{\partial ^2 H}{\partial x^2}-2V^3r^4(\frac{\partial H}{\partial x})^2-3VH^2r^4\frac{\partial ^2 V}{\partial r^2}\nn\\
%&+&2V^2r^4H\frac{\partial ^2 H}{\partial r^2}-2V^2r^4(\frac{\partial H}{\partial r})^2+3H^2r^4(\frac{\partial V}{\partial r})^2+4V^2H
%r^3\frac{\partial H}{\partial r}-6VH^2r^3\frac{\partial V}{\partial r}-3n^2H^2\nn\\
%&+&3\Lambda H^{8/3}V^3r^4+2V^3r^4(\frac{\partial H}{\partial x})^2+2V^2r^4(\frac{\partial H}{\partial r})^2=0.
%\label{Geq}
%\end{eqnarray} 
The partial differential equation (\ref{A1}) leads to considering the 
separation of variables in the metric function, according to
\be
H(t,r,x)=H_1(r)H_2(x)+H_3(t)\label{sep}.
\ee
After substitution (\ref{sep}) for the metric function $H(t,r,x)$ in the other differential equations (\ref{A2})-(\ref{A9}), we find, quite interestingly, that $H_1(r)$ and $H_2(x)$ satisfy the same differential equations  (\ref{Rde}) and (\ref{XDeq}) for $R(r)$ and $X(x)$, respectively.  Moreover, we find that the solutions for $H_3(t)$ are given by 
\be
H_3(t)=\epsilon+\rho_D t
\ee
where $\rho_D={\frac{D-3}{\ell}}$ in which the $D$-dimensional length scale $\ell$ is related to the $D$-dimensional cosmological constant  $\Lambda$ by $\ell=\sqrt{\frac{(D-1)(D-2)}{2\Lambda}}$. We also note that in the limit of $\Lambda \rightarrow 0$, the general solutions based on the metric function (\ref{sep}) should reduce to the previous solutions, with the metric function (\ref{HGEH}). This requirement implies that $\epsilon=1$. As a result, we find that the most general solution for the metric function $H(t,r,x)$ is
\be
H(t,r,x)=1+ \rho_D t+\frac{\chi}{2\xi_D}\int _{0}^\infty dc\, {x}^{\frac{7-D}{2}}\,c^{\frac{D-1}{2}}{ H_C} \left( 0,0,0,-\frac{{
a}^{2}{c}^{2}}{2},\frac{{a}^{2}{c}^{2}}{4},{\frac {{a}^{2}-{r}^{2}}{2{a}^{
2}}} \right)
{{K_{\frac{D-7}{2}}}\left(cx\right)} \label{HFINCOS}.
%\end{eqnarray}
\ee
We shall emphasis that in the limit of $a\rightarrow 0$, the metric function (\ref{HFINCOS}) reduces exactly to the metric function $H_0(t,r,x)$ of the following 
exact $D$-dimensional solutions to the Einstein-Maxwell theory with the cosmological constant $\Lambda$
\be
ds^2=-\frac{1}{H_0^2(t,r,x)}dt^2+H_0(t,r,x)^{2/(D-3)}(dx^2+dw^2+w^2d\Omega_3^2).
\ee
The metric function $H_0(t,r,x)$ is given by 
\be
H_0(t,r,x)=1+\rho_D t+\frac{\chi}{(r^2+x^2)^{\frac{D-3}{2}}} \label{H0},
\ee
and 
\be
{A_t}(t,r,x)={\sqrt{\frac{D-2}{D-3}}H_0^{-1}(t,r,x)}\label{gaugeD0time}.
\ee
%We note that 
 %(\ref{Ehazero}), we find that in presence of cosmological constant,  we find the exact solutions 
%to the Einstein-Maxwell theory along with the gauge field (\ref{gaugeD}) where the metric function $H_0(r,x)$ is given by
Moreover, we find the second class of cosmological solutions that correspond to solutions (\ref{GENHGEH2}) and are given by
%Similar calculations show in $D$ dimensions, we get two general cosmological convoluted solutions, given by
\begin{equation}
\tilde H(t,r,x)=1+ \rho_D t+\frac{\chi \pi  \epsilon_D}{4\xi_D}\int _{0}^\infty dc\, {c^{\frac{D-1}{2}}}\, {x}^{\frac{7-D}{2}}\,{ { H}_C} \left( 0,0,0,\frac{{a}^{2}{c}^{2}}{2},-\frac{{a}^{2}{c}^{2}}{4},{\frac {{a}^{2}-{r}^{2}}{2{a}^{
2}}} \right)\,
{{J_{\frac{D-7}{2}}}\left(cx\right)}
 \label{GENHGEH2COS}.
\end{equation}
We also notice that in asymptotic region $r \rightarrow 0$, the cosmological metric (\ref{mD}) approaches to
%\be
%ds_2=(1+\rho_D t)^{-2}dt^2+(1+\rho_D t)^{\frac{2}{D-3}}(dx^2+x^2d\Omega_{D-6}+ds_{EH}^2)
%\ee
%or
\be
ds_D^2=-dT^2+e^{\frac{2\rho_D}{D-3}T}(dx^2+x^2d\Omega_{D-6}+ds_{EH}^2)
\ee
where $T=\frac{1}{\rho_D}\ln (1+\rho_D t)$. %, $dT^2=\frac{dt^2}{(1+\rho_D t)^2}$
The constant-$T$ hypersurfaces have the smallest hypervolume at $T=0$ and for later times, the hypervolume increases exponentially with time. 
Moreover, the cosmological
%Since the general solutions (\ref{HTNDcos}) and (\ref{HTNDsecondcos}) describe the asymptotically dS spacetime, we 
%take a look at the cosmological 
$c$-function for asymptotically dS solutions (\ref{mD}) with the metric function (\ref{HFINCOS}) and (\ref{GENHGEH2COS}), that is defined by
%For any %asymptotically $D$-dimensional dS spacetime, one can define the $c$-function 
\cite{Lob}
\be
c_D\simeq \frac{1}{\left( G_{\mu \nu }n^{\mu }n^{\nu }\right) ^{\frac{D}{2}-1}},\label{cf}
\ee
shows that the $c$-function is an increasing function of time. In equation (\ref{cf}), $n^{\mu }$ is a temporal unit vector.% pointing toward the time direction. 

\section{Concluding remarks}
\label{sec:con}
In this article, we construct exact solutions in Einstein-Maxwell theory in six and higher dimensions in which Eguchi-Hanson instanton is the base space. We show that all field equations reduce to a partial differential equation for the metric function. We analytically solve the partial differential equation in any dimensions and find the most general solutions for the metric function. The exact solutions are given by the metric functions (\ref{HFIN}) and (\ref{GENHGEH2}) where we have found the proper measure functions by comparing the solutions to some exact analytical solutions in Einstein-Maxwell theory (equations (\ref{Ehazero}) and (\ref{H09})) in some proper limits. 

 %(%) and (). sp which the metric function depends on integral of two convoluted-like functions of the non-compact coordinates. The two classes of solutions are given by the line element (\ref{mD}) and the electromagnetic gauge field (\ref{gaugeD}) in which the metric function $H(r,x)$ for two different classes of sloutions are given by  respectively. In deriving the metric functions, we consider some proper limits of the solutions and compare them to the exact analytical solutions in $D$-dimensions that are given by   

%Except on the location of a bolt at the origin of spherical coordinates, all the metric functions are regular in any points of spacetime. The bolt singularity may be converted to a higher dimensional regular hypersurface if we consider the metric functions that depend on more than two spatial directions. 

We also find similar solutions to the Einstein-Maxwell theory in presence of cosmological term. In the solutions, the metric functions not only depend on two spatial directions, but also depends on the time coordinate. We find that in different dimensions, all Einstein's and Maxwell's equations lead to a partial differential equation for the metric function. We use a special separation of variables to solve the partial differential equation in two spatial directions and the time coordinate.  These exact solutions are given by the metric functions (\ref{HFINCOS}) and (\ref{GENHGEH2COS}), that linearly depends on time.  %The solutions in asymptotic region n by the metric (\ref{mD}) and gauge field (\ref{gaugeD}) and the cosmological metric functions (\ref{HTNDcos}) and (\ref{HTNDsecondcos}). 
We also find that asymptotically, the cosmological solutions for a fixed time slice, represent the expanding patches of dS spactime, in agreement with the monotonic increasing $c$-functions.% in agreement with the $c$-theorem for asymptotically dS spacetimes.

\bigskip
{\Large Acknowledgments}

This work was supported by the Natural Sciences and Engineering Research
Council of Canada.

\appendix

\section{The equations for the time dependent metric function $H(t,r,x)$ in different dimensions}
\label{Ap1}
The Einstein's equations imply
\be
\frac{\partial ^2 H(t,r,x)}{\partial x\partial t}=\frac{\partial ^2 H(t,r,x)}{\partial r\partial t}=0,\label{A1}
\ee
as well as the following equations in different dimensions:
In six dimensions, we find
\begin{eqnarray}
&-&12 H \left( t,r,x \right) ^{11/3} {\frac {\partial ^{2}}{\partial {t}^{2}}}H \left( t,r,x
 \right)  {r}^{
5}-20   H \left( t,r,x \right) ^{8/3}  {\frac {
\partial }{\partial t}}H \left( t,r,x  \right) ^{2}{r}^{5}+18
 \Lambda H \left( t,r,x  \right) ^{8/3}{r}^{5}\nn\\
&-&12 H \left( t,r,x \right) 
  {\frac {\partial ^{2}}{\partial {r}^{2}}}H \left( t,r,x
 \right) {a}^{4}r+12  H
 \left( t,r,x \right) {\frac {
\partial ^{2}}{\partial {r}^{2}}}H \left( t,r,x \right)  {r}^{5}+12  H \left( t,r,x
 \right) {\frac {\partial ^{2}}{
\partial {x}^{2}}}H \left( t,r,x \right)    {r}^{5}\nn\\
&+&12 H \left( t,r,x \right)  {\frac {\partial }
{\partial r}}H \left( t,r,x   \right) {a}^{4}+36 H \left( t,r,
x \right)   {\frac {\partial }{\partial r}}H \left( t,r,x
 \right)  {r}^{4}=0,\label{A2}
\end{eqnarray}
and in seven dimensions
\begin{eqnarray}
&-&10  H \left( t,r,x \right) ^{7/2} {\frac {\partial ^{2}}{\partial {t}^{2}}}H \left( t,r,x
 \right)  {r}^{5
}x-15  H \left( t,r,x \right)   ^{5/2} {\frac {
\partial }{\partial t}}H \left( t,r,x \right)  ^{2}{r}^{5}x+16
 \Lambda  H \left( t,r,x  \right) ^{5/2}{r}^{5}x\nn\\
&-&10
  H \left( t,r,x \right){\frac {\partial ^{2}}{\partial {r}^{2}}}H \left( t,r,x
 \right)   {a}^{4}rx+10 H
 \left( t,r,x \right) {\frac 
{\partial ^{2}}{\partial {r}^{2}}}H \left( t,r,x \right)  {r}^{5}x+10H \left( t,r,x
 \right)  {\frac {\partial ^{2}}{
\partial {x}^{2}}}H \left( t,r,x  \right)  {r}^{5}x\nn\\
&+&10 H \left( t,r,x \right)   {\frac {\partial 
}{\partial r}}H \left( t,r,x \right) {a}^{4}x+30 H \left( t,
r,x \right)  {\frac {\partial }{\partial r}}H \left( t,r,x
 \right)  {r}^{4}x+10 H \left( t,r,x \right)  {\frac {\partial }{\partial x}}
H \left( t,r,x \right)  {r}^{5}=0.\nn\\
&&
\end{eqnarray}
Moreover, in eight dimensions, we find
\begin{eqnarray}
&-&30 H \left( t,r,x   \right) ^{{\frac {17
}{5}}}  {\frac {\partial ^{2}}{\partial {t}^{2}}}H \left( t,r,x
  \right)  {r}^{5}x-42  H \left( t,r,x \right) ^{{\frac {
12}{5}}} \left({\frac {\partial }{\partial t}}H \left( t,r,x 
 \right) \right)^{2}{r}^{5}x-30 H \left( t,r,x \right)  {\frac {\partial ^{2}}{\partial {r}^
{2}}}H \left( t,r,x  \right)  {a}^{4}rx\nn\\
&+&
30H \left( t,r,x \right)    {\frac {\partial ^{2}}{\partial {r}^{2}}}H \left( t,r,x
 \right)  {r}^{5}x+30  H
 \left( t,r,x \right) {\frac {
\partial ^{2}}{\partial {x}^{2}}}H \left( t,r,x   \right) {r}^{5}x+30 H \left( t,r,x \right)  {
\frac {\partial }{\partial r}}H \left( t,r,x \right)  {a}^{4}x\nn\\
&+&90 H \left( t,r,x \right)  {\frac {\partial }{\partial r}}H
 \left( t,r,x \right)   {r}^{4}x+60 H \left( t,r,x \right)  {\frac {\partial }
{\partial x}}H \left( t,r,x  \right)  {
}^{5}+50 \Lambda   H \left( t,r,x \right) ^{{\frac {
12}{5}}}{r}^{5}x=0,
\end{eqnarray}

and correspondingly in nine dimensions, we get 

\begin{eqnarray}
&-&21H \left( t,r,x  \right) ^{10/3}   {\frac {\partial ^{2}}{\partial {t}^{2}}}H \left( t,r,x
 \right)  {r}^{
5}x-28   H \left( t,r,x   \right) ^{7/3} \left( {\frac {
\partial }{\partial t}}H \left( t,r,x \right)  \right) ^{2}{r}^{5}x+36
 \Lambda  H \left( t,r,x \right) ^{7/3}{r}^{5}x\nn\\
&-&21
 H \left( t,r,x \right)  {\frac {\partial ^{2}}{\partial {r}^{2}}}H \left( t,r,x
 \right)  {a}^{4}rx+21 H
 \left( t,r,x \right)   {\frac 
{\partial ^{2}}{\partial {r}^{2}}}H \left( t,r,x  \right) {r}^{5}x+21 H \left( t,r,x
 \right)  {\frac {\partial ^{2}}{
\partial {x}^{2}}}H \left( t,r,x  \right)  {r}^{5}x\nn\\
&+&21 H \left( t,r,x \right)  {\frac {\partial 
}{\partial r}}H \left( t,r,x  \right) {a}^{4}x+63 H \left( t,
r,x \right)  {\frac {\partial }{\partial r}}H \left( t,r,x
 \right)  {r}^{4}x+63  H \left( t,r,x \right){\frac {\partial }{\partial x}}
H \left( t,r,x  \right)  {r}^{5}=0.\nn\\
&&
\end{eqnarray}

In six dimensions, the Maxwell equations imply the partial differential equation

\be
 \left( {\frac {\partial ^{2}}{\partial {r}^{2}}}H \left( t,r,x
 \right)  \right) ({a}^{4}r-r^5) - \left( {\frac {
\partial ^{2}}{\partial {x}^{2}}}H \left( t,r,x \right)  \right) {r}^{
5}- \left( {\frac {\partial }{\partial r}}H \left( t,r,x \right) 
 \right) {a}^{4}-3  \left( {\frac {\partial }{\partial r}}H \left( t,
r,x \right)  \right) {r}^{4}=0.
\ee
In seven dimensions, we find
\begin{eqnarray}
&& \left( {\frac {\partial ^{2}}{\partial {x}^{2}}}H \left( t,r,x
 \right)  \right) {r}^{5}x- \left( {\frac {\partial ^{2}}{\partial {r}
^{2}}}H \left( t,r,x \right)  \right) {a}^{4}rx+ \left( {\frac {
\partial ^{2}}{\partial {r}^{2}}}H \left( t,r,x \right)  \right) {r}^{
5}x+ \left( {\frac {\partial }{\partial r}}H \left( t,r,x \right) 
 \right) {a}^{4}x\nn\\
&+&3  \left( {\frac {\partial }{\partial r}}H \left( t
,r,x \right)  \right) {r}^{4}x+ \left( {\frac {\partial }{\partial x}}
H \left( t,r,x \right)  \right) {r}^{5}=0,
\end{eqnarray}
and in eight dimensions, we get the equation
\begin{eqnarray}
 &&\left( {\frac {\partial ^{2}}{\partial {x}^{2}}}H \left( t,r,x
 \right)  \right) {r}^{5}x- \left( {\frac {\partial ^{2}}{\partial {r}
^{2}}}H \left( t,r,x \right)  \right) {a}^{4}rx+ \left( {\frac {
\partial ^{2}}{\partial {r}^{2}}}H \left( t,r,x \right)  \right) {r}^{
5}x+ \left( {\frac {\partial }{\partial r}}H \left( t,r,x \right) 
 \right) {a}^{4}x\nn\\
&+&3  \left( {\frac {\partial }{\partial r}}H \left( t
,r,x \right)  \right) {r}^{4}x+2  \left( {\frac {\partial }{\partial 
x}}H \left( t,r,x \right)  \right) {r}^{5}=0.
\end{eqnarray}
Finally, in nine dimensions, we find
\begin{eqnarray}
&& \left( {\frac {\partial ^{2}}{\partial {x}^{2}}}H \left( t,r,x
 \right)  \right) {r}^{5}x- \left( {\frac {\partial ^{2}}{\partial {r}
^{2}}}H \left( t,r,x \right)  \right) {a}^{4}rx+ \left( {\frac {
\partial ^{2}}{\partial {r}^{2}}}H \left( t,r,x \right)  \right) {r}^{
5}x+ \left( {\frac {\partial }{\partial r}}H \left( t,r,x \right) 
 \right) {a}^{4}x\nn\\
&+&3  \left( {\frac {\partial }{\partial r}}H \left( t
,r,x \right)  \right) {r}^{4}x+3  \left( {\frac {\partial }{\partial 
x}}H \left( t,r,x \right)  \right) {r}^{5}=0.\label{A9}
\end{eqnarray}

\end{document}